\documentstyle[floats,aps,epsf,prl]{revtex}

\begin{document}
\draft
\preprint{October 14, 2002}

\twocolumn[\hsize\textwidth\columnwidth\hsize\csname
@twocolumnfalse\endcsname

\title{Anyon Wave Function
for the Fractional Quantum Hall Effect}
\author{Orion Ciftja,$^{1}$ George S. Japaridze,$^{2}$ 
and Xiao-Qian Wang$^{2}$}
\address
{$^{1}$Department of Physics, Prairie View A\&M University, 
Prairie View, Texas 77446 \\
$^{2}$Department of Physics, 
Clark Atlanta University, Atlanta,
Georgia 30314}
\date{\today}
\maketitle

\begin{abstract}
An anyon wave function 
(characterized by the statistical factor $n$) projected
onto the lowest Landau level is derived for    
the fractional quantum Hall effect states at filling factor
$\nu = n/(2pn+1)$ ($p$ and $n$ are integers).
We study the properties of the anyon wave function by using
detailed Monte Carlo simulations in disk geometry and show that
the anyon ground-state energy is a lower bound to the composite
fermion one.
%
%The ground state energies for the anyon wave functions 
%are calculated using Monte Carlo  
%simulations and are shown to be the lower bound 
%for the composite fermions.
%
Our results suggest that the composite fermions
can be viewed as a combination of anyons and a
fluid of charge-neutral dipoles.
\end{abstract} 


\vskip0pc]

The fractional quantized Hall effect (FQHE) is one
of the most fascinating phenomena in condensed-matter physics~\cite{book}.
The pioneering work by Laughlin~\cite{Laughlin} based on the famous 
trial wave function
at filling of $\nu= 1/(2p+1)$
revealed that FQHE arises from the
formation of an incompressible quantum fluid
that supports quasiparticles and quasiholes carrying fractional
charge and statistics.
The Jain's composite fermion (CF) approach~\cite{Jain}
successfully clarified fundamental aspects
of the FQHE, which evolved into the description of the 
FQHE in terms of electron-vortex composites. 
A CF is the bound state of an electron and an even number of vortices 
formed in a two-dimensional (2D) system of electrons subject to a strong 
perpendicular magnetic field. 
Based on the CF theory, the interacting electrons at the 
Landau level (LL) filling factor $\nu = n/(2pn+1)$, 
$n$ and $p$ being integers,
transform into weakly interacting CFs with
an effective filling factor $\nu^{\star}=n$, corresponding to 
$n$-filled CF LLs.

The connection between the FQHE and the integer quantum Hall effect (IQHE)
has motivated 
the Chern-Simons (CS) field theoretical approach~\cite{LF,HLR}
for the FQHE. Within this field theoretical approach, an even 
number of magnetic flux quanta 
($\phi_{0} = hc/e$ stands for one flux quantum) 
are attached to the 2D electrons through the introduction of
a 
CS gauge field. 
In a mean-field approximation where the statistical gauge
fluxes are delocalized from the electrons and uniformly spread out
in the 2D plane, the average CS gauge field partially cancels the
external magnetic field. So far, the fermion CS approach has
been very successful to describe the nature of the quantum Hall state
at $\nu = 1/2$ where 
the CS flux generated fictitious magnetic field cancels exactly 
the external magnetic field at the mean-field level.    
 
An important aspect for developing theories of FQHE is that
the Hilbert space is composed solely of states in 
the lowest Landau level (LLL). This constraint is important conceptually, 
as it implies the quenching of the kinetic energy and  
the non-perturbative nature of electron-electron interactions. 
Because of the drastic reduction of the size of the Hilbert space,  
numerical studies have been indispensable 
to the development of our understanding of the FQHE. 
Although the FQHE can be considered as an
IQHE of CFs, it is worth noting that 
an explicit LLL projection of the latter states is necessary in the  
construction of Jain's wave function in order to obtain
the correct low-energy physics~\cite{Jain}.

The Chern-Simons field theory is intimately connected to 
the statistics transmutation of anyons. Halperin~\cite{Hal} 
demonstrated that an anyon picture inherently leads to an 
explanation of the 
quantum Hall hierarchy. 
Similarly, Ma and Zhang~\cite{MaZhang} considered ideal anyons subject
to a magnetic field.   
The ideal anyons with statistics $1/n$ in a strong magnetic field 
have a ground state that exhibits an IQHE at filling factor $n$ 
with quasiparticle excitations of charge $n e$. 
The electron FQHE states are realized
with asymmetry in quasiparticle states ($\nu= n/(2pn+1)$) and quasihole
states ($\nu= n/(2pn-1)$) in the fractional
quantum Hall hierarchy. Despite the consensus that anyons 
play an important role in understanding the
FQHE, there remains no explicit, quantitative study
of the corresponding ground-state properties.    

In this paper we 
derive an
anyon wave function for filling factors $n/(2pn+1)$ with $p$ and $n$ 
being integers that is fully projected onto the LLL.
We then study the properties of the anyon wave function at these
filling factors by carrying out  
detailed Monte Carlo (MC) simulations.
Our calculation provides important information on the
connection between anyons and CFs. 

{\it The Chern-Simons transformation.}$-$
The Hamiltonian for 2D electrons subject to the perpendicular magnetic field,
${\bf B}= -B {\bf \hat{z}}$ is given by 
\begin{equation}
{\hat H} = {\hat H}_{0} (B) +\hat{V} = \frac{1}{2m} \sum_{j=1}^{N} ({\bf p}_{j} 
+ e {\bf A}_{j})^{2} + {\hat V}, 
\end{equation}
where $m$ and $-e$ are electron's mass
and charge, $N$ is the total number of electrons
and $\hat{V}$ is the Coulomb interaction energy.
We adopt the symmetric gauge in which the vector 
potential is given by ${\bf A}_{j} =(B/2)(y_{j}, -x_{j}, 0)$. 
With the use of complex coordinates $z_{j}=x_{j}+iy_{j}$, and the 
notation $\chi = \prod_{i<j}^{N}(z_{i}-z_{j})$,  
the CS transformation amounts to 
multiplying the many-body wave function of the CS transformed Hamiltonian
with 
 $(\chi / | \chi |)^{\alpha}$ 
and 
$e{\bf A}_j\rightarrow e{\bf A}_j+\alpha {\bf {\cal A}}_j\equiv e{\bf A}_j +
\alpha\sum^N_{k\neq j}(-(y_{j}-y_k),x_{j}-x_k,0)/ |{\bf r}_j - {\bf r}_k|^2$.

After the CS transformation, the kinetic energy operator
${\hat H}_{0} (\alpha)$  can be expressed 
in terms of destruction and creation operators, $a(\alpha)$ and 
$a^\dagger(\alpha)$, as 
(all lengths are measured in units of magnetic length $l\equiv \sqrt{\hbar 
/e|B|}$):
\begin{equation}
\label{cs4}
{\hat H}_{0}(\alpha)=
\hbar \omega_c\sum_{j=1}^{N} \,a^\dagger_j(\alpha)\,a_j(\alpha)\,+\,
\frac{1}{2} N \hbar \omega_{c},
\end{equation}
where  $\omega_{c}=e|B|/m$ is the cyclotron frequency and 
\begin{eqnarray}
a_j(\alpha) & = & \frac{1}{\sqrt{2}} 
(2\frac{\partial}{\partial z^\star_j}\,+\,
\frac{z_j}{2}+\alpha\frac{\partial \ln\chi^\star}{\partial z^\star_j}),
 \\
a^\dagger_j(\alpha) & = & \frac{1}{\sqrt{2}} (-2\frac{\partial}{\partial
z_j}+\frac{z_{j}^{\star}}{2}+\alpha\frac{\partial \ln\chi}{\partial
z_j} ).
\end{eqnarray}
The operators 
$a^\dagger_j(\alpha)$ and $a_j(\alpha)$ satisfy Bose 
commutation relations, $[a_j(\alpha), a^{\dagger}_k(\alpha)]=\delta_{jk}$. 

The angular momentum operator ${\hat L}(\alpha)$ after the transformation 
reads
\begin{equation}
\label{L}
{\hat L} \rightarrow {\hat L}(\alpha)=\sum^N_{j=1}( b^\dag_j(\alpha)b_j(\alpha)- 
a^\dag_j(\alpha)\,a_j(\alpha) ),
\end{equation}
where the operators $b(\alpha)$ and $b^\dag(\alpha)$ describe
the LLL degeneracy, 
\begin{eqnarray}
 b_j(\alpha) & = &  \frac{1}{\sqrt{2}}( 2\frac{\partial}{\partial z_j}
+\frac{z^\star_j}{2} 
-\alpha \frac{\partial \ln \chi}{\partial z_j}), \\
 b^\dag_j(\alpha) & = & \frac{1}{\sqrt{2}}
( -2\frac{\partial}{\partial z^\star_j}
+\frac{z_j}{2}  
-\alpha \frac{\partial \ln 
\chi^\star}{\partial z^\star_j}),
\end{eqnarray}
satisfying $[b_j(\alpha), b^{\dagger}_k(\alpha)]=\delta_{jk}$.

So far, two types of CS 
transformations have been discussed 
in the literature. 
One considers $\alpha$ to be an integer ($\alpha=2p$ has been used in the fermion 
CS theory to describe the reduction of the effective 
magnetic field felt by the CFs~\cite{LF,SH,HSH}; and $\alpha$ being an odd integer has been used
to describe the formation of Bose condensate in FQHE~\cite{bn}). 
The other choice concerns $\alpha = 1/n$ and has been used for the study of 
anyon superconductivity in the absence of a magnetic field~\cite{fhl}.

We recall that for $\hat{H}_0(B)$, the LLL wave function has the form
 $\Psi_{LLL}(B)=W(N)\, f(\{ z_{j} \},\{ \partial / \partial  z_{j} \} ) $,  
where $W(N)=\exp (-\sum_{i=1}^{N} |z_{i}|^{2}/4 l^2)$~\cite{Jain}.
%
%As the consequence 
%of the $\cal {PT}$-invariance of the Hamiltonian, 
%
Due to the symmetry properties 
of parity ($\cal P$) and time-reversal ($\cal T$) of the Hamiltonian 
(under   
${\cal PT}$: $z\leftrightarrow -z^\star$, $B\leftrightarrow 
-B$, $\alpha\leftrightarrow -\alpha$), 
the corresponding wave function for 
${\hat H}_{0} (-B)$ is $\Psi_{LLL}(-B)=W(N)\,f(\{ z_{j}^{\star} \},
\{ \partial / \partial  z_{j}^{\star} \} )$. 
$\Psi_{LLL}(B)$ and $\Psi_{LLL}(-B)$ 
are the eigenfunctions 
for ${\hat L}(\alpha=0)$ as well, with the eigenvalues $N (N-1)/2\nu$ and
$- N (N-1)/2\nu$ ($\nu$ is the filling factor), respectively.  
Furthermore,  
$f$ should be an antisymmetric function for electrons.

One can get useful information on the explicit form of $f$ 
by employing the CS transformation 
and the $\cal {PT}$ symmetry property of the LLL wave functions.
For our purpose, we are interested in deriving a general equation for the 
LLL projection by requesting 
\begin{equation}
\label{grstate}
a_j(\alpha)\Psi_{LLL}=0.
\end{equation}
From Eqs. (3) and (8), we have:
\begin{equation}
\label{f}
2\frac{
\partial  \ln f}{\partial z^\star_{j}}+
\alpha
\frac{\partial \ln \chi^\star}{\partial z^\star_{j}}=0.
\end{equation}

Note that for $f(\{ z_{j} \},\{ \partial / \partial  z_{j} \} )$, 
Eq. (9) yields
a trivial solution $\alpha=0$. 
For $f(\{ z_{j}^{\star} \},
\{ \partial / \partial  z_{j}^{\star} \} )$, Eq. (9) is
an operator equation and the solution is a functional of $\chi^{\star}$. 
If we approximate $f(\{ z_{i}^{\star} \},\{ \partial / \partial  z_{i}^{\star} \} )$
by a function form $f(\{ z_{i}^{\star} \})$, 
there exists a simple solution:
\begin{equation}
\label{fsol}
f(\{ z^\star_{j} \})=(\chi^\ast)^{-\alpha/2}.
\end{equation}
Furthermore, 
using the constraint that $\Psi_{LLL}(-B)$ is the eigenfunction of 
${\hat L}(\alpha)$ with the eigenvalue $N(N-1)/2 \nu$,  
one determines $\alpha$ in terms of the filling factor
$\alpha=-2/\nu$.

By putting everything together, we arrive at the LLL wave function for 
fillings of $\nu =n/(2pn+1)$:
\begin{equation}
\Psi_{LLL} (B)=\prod_{i<j}^{N} (z_i-z_j)^{2 p+\frac{1}{n}} 
\exp(-\sum_{i=1}^{N} |z_i|^{2} / 4l^{2}) .
\label{laughlin}
\end{equation}

A few remarks are immediately in order. 
(i) For $n=1$, Eq. (11) coincides with Laughlin's 
    wave function for $\nu = 1/(2p+1)$.
   For $n>1$,  the wave function describes an
incompressible quantum liquid state with   
    filling factor $\nu = n/(2pn+1)$ ($p$ and $n$ are integers) and
    is not antisymmetric (as requested for a fermion state).
    It has anyon symmetry (characterized by the statistical 
parameter $n$). The LLL anyon wave function is expected to
    serve as a lower bound for the energy of the electronic system. 
Eq. (11) describes composite anyons 
in that the anyons are attached by $2p$ vortices.  
(ii) The transformation employed is a unitary one, in contrast 
     to that used by Ma and Zhang~\cite{MaZhang}. 
     As a result, our approach conserves the density of the
     system upon transformation. Nevertheless it should be noted that the
     resulting wave functions are of the same form. The quasiparticle and quasiholes 
obey fractional statistics and have fractional charge~\cite{MaZhang}.
     Therefore, there exists an asymmetry 
for the quasiparticle and quasihole
states. The quasiparticle 
states ($\nu=n/(2pn+1)$) correspond to anyon screening (i.e., the 
anyon flux generated fictitious 
magnetic field cancels the residual external fields), 
while the quasihole 
states ($\nu=n/(2pn-1)$) correspond to anyon anti-screening (for the anyon equation of state in the anti-screening regime, see ~\cite{ouvry}).  
(iii) The average CS-generated magnetic field reverses
      the external magnetic field. 
      This can be seen to be in agreement to the CS 
transformations for CFs ($\alpha=2p$) for the reduction of the effective magnetic fields for the CFs. For the part associated with the anyons, the
coefficient $2/n$ in the CS transform  
can be understood as follows: $1/n$-CS field generates
a fictitious magnetic field
that cancels the external magnetic field, while the additional $1/n$-CS
reverses the direction of the external magnetic field with
the same magnitude that corresponds to $n$-filled Landau bands. 
     It becomes clear that after expansion of the anyon's wave function in the 
fermion representation where the average-field 
corresponds to $n$-filled Landau bands, 
the resulting wave function after the LLL projection 
becomes Jain's CF wave function. 
It is worth pointing out that the cancellation or the effective reduction of magnetic field
is a dynamical feature 
associated with the CS gauge field.
(iv) According to the corresponding state theory of the global phase diagram~\cite{lkz}, 
the incompressible quantum fluid states described by the anyon wave functions correspond to
superfluid states. For FQHE, the transition between different 
integer quantum Hall states is predicted to follow a
floating-up picture, i.e., only transitions between adjacent $n$ are allowed. 
In the floating-up scenario, one of us~\cite{xqw} argued that  
the integer quantum Hall transitions
correspond to the transition from superfluid state 
for anyons characterized by $n$ ($n$-anyons) to insulating states of
$(n\pm 1)$-anyons.  
This is in agreement with the hierarchy picture for Eq. (11).  

{\it Monte Carlo calculations.}$-$It is readily observable that the 
systems described by the anyon wave function can be mapped
to a one-component plasma with inverse temperature $(4p+2/n)$~\cite{Laughlin}. 
%To obtain quantities of interest, we performed standard MC simulations 
%in disk geometry. 
The system under consideration consists of $N$ particles
moving in a 2D space subject to a strong
perpendicular magnetic field and embedded in a uniform  neutralizing
background of positive charge.
Our goal is to calculate the thermodynamic limit of the expectation 
value of the potential energy operator and other quantities.
To do so we perform detailed MC simulations in disk
geometry~\cite{morf} and extract the thermodynamic estimate of 
various quantities by extrapolating the finite $N$ results.
 
In our simulations we adopt the well-known 
Metropolis algorithm \cite{metropolis}.
The expectation value of any operator is then estimated by averaging
its value over numerous configurations. 
For each $N$ we routinely perform MC simulations employing several
million configurations.
All the results that we report here were obtained after discarding 100,000 
``equilibration'' MC steps and using 
$2 \times 10^6$ MC steps for averaging purposes.                      
The thermodynamic estimate of the correlation energy per particle
is obtained by fitting $E=\langle \hat{V} \rangle/N$ with a second-order
polynomial in $1/\sqrt{N}$ by using systems with
$N=4,16,36,64,100,144$, and $196$ particles.

In Table~\ref{tabstandard} we show the thermodynamic estimates
of the correlation energy per particle for a number of FQHE states
described by the anyon wave function.
The results are rounded in the last digit.
We compare the energies obtained from the anyon wave function with the
corresponding results of the LLL projected Jain's CF wave function
as obtained after MC simulations in spherical geometry~\cite{ejain}.

\begin{table}
\caption[]{Thermodynamic estimates for the correlation energy per particle 
           of various FQHE states described by the anyon and
           Jain's CF wave function~\cite{Jain}.
           Energies are in units of $e^2/l$.}
\label{tabstandard}
\begin{center}
\begin{tabular}{ccccc}
 $p$ & $n$ & $\nu$     & $E$  & $E_{CF}$    \\ \hline
 1 & 1 & 1/3          & -0.4095   & -0.4098  \\          
   & 2 & 2/5          & -0.4422   & -0.4328   \\     
   & 3 & 3/7          & -0.4550   & -0.4423    \\    
   & 4 & 4/9          & -0.4618   & -0.4474     \\   
   & 5 & 5/11         & -0.4661   & -0.4508      \\  
   & $\infty$ & 1/2          & -0.4844   & -0.4653 \\       
 2 & 1 & 1/5   & -0.3274 & -0.3275   \\ 
   & 2 & 2/9   & -0.3432 & -0.3428   \\
   & 3 & 3/13  & -0.3490 & -0.3483   \\ 
   & 4 & 4/17  & -0.3520 & -0.3512    \\
   & 5 & 5/21  & -0.3538 & $-$    \\
   & $\infty$ & 1/4          & -0.3615   & $-$
\end{tabular}
\end{center}
\end{table}

The energy difference between Jain's CF and anyon states can be 
interpreted
as an exchange correction in that the anyon wave function
does not satisfy the Pauli principle requested for fermions 
(except for $n=1$).
As can be seen from Table~\ref{tabstandard},
the energy of the anyon state, $E$, serves as a lower bound 
for the CF's energy $E_{CF}$.
It is interesting to note that such an exchange correction, $E_{X}$,
satisfies the relation 
\begin{equation}
E_{X}= E_{CF} - E = (1 - 1/n) \ \Delta_{p},
\end{equation} 
where $\Delta_{p}$ can be readily extracted from Table~\ref{tabstandard}.
Our calculations show that
 $\Delta_{1} = 0.019 \ e^2/l$ and $\Delta_{2} \simeq 0.001 \ e^2/l$.  

The 
$n$-dependence of the exchange correction is of the same form, $|1-1/n|$, as 
Ma and Zhang used in the perturbative
analysis of ideal anyons in a magnetic field~\cite{MaZhang}.  
The size of the exchange correction indicates that the
anyon wave function captures the essential physics of the FQHE. 
The largest correction refers to the case of $\nu=1/2$, where
$E$ is merely 4\% lower than $E_{CF}$.  
The effect of exchange correction on the energy is 
greatly reduced with increase of $p$. In fact, for $p>1$, 
the contribution of the exchange correction becomes negligible.
 
Following the same procedure as described in Ref.~\cite{morf},
we have calculated the pair distribution, $g(r)$.  
As seen from Fig. 1, the calculated
$g(r)$ shows $p~(=1,2)$ ``bumps" for filling factors of $n/(2pn+1)$. 
This is to be compared with the results for CFs, where more 
``bumps" (or ``wiggles'') are observable due to the inherent
fermion symmetry of Jain's CF wave function and the associated 
Friedel-like
oscillations.
Moreover, there exists notable differences in the short distance behavior
of $g(r)$. 
The short-range behavior of the pair distribution
function obtained from the anyon wave function is consistent with
$g(r) \propto r^{4p+2/n}$.
At $\nu=1/2$, the $r^{4}$-dependence is  
to be compared with the 
$g(r) \propto r^{2}$ dependence obtained from 
Rezayi-Read Fermi wave function~\cite{rezayi}.

\begin{figure}
\begin{center}
\epsfxsize=3.2in
\epsffile{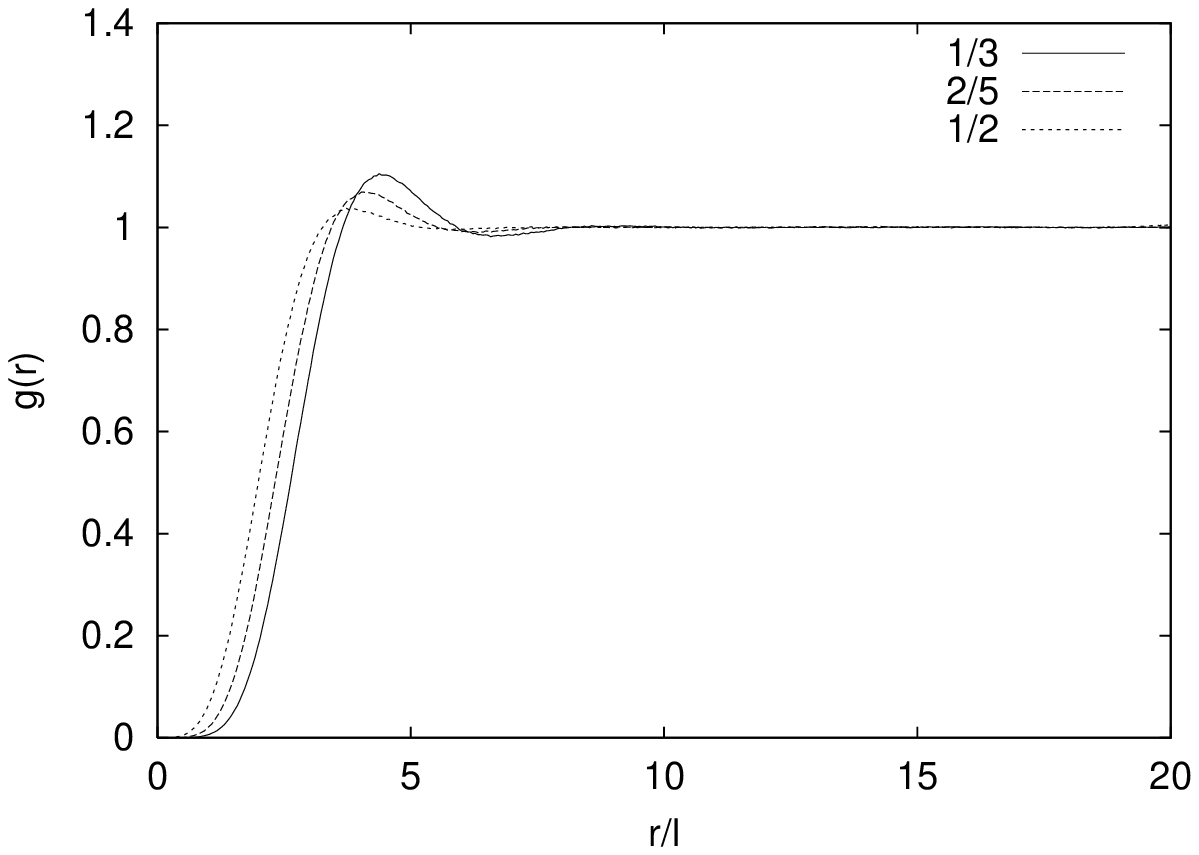}
\epsfxsize=3.2in
\epsffile{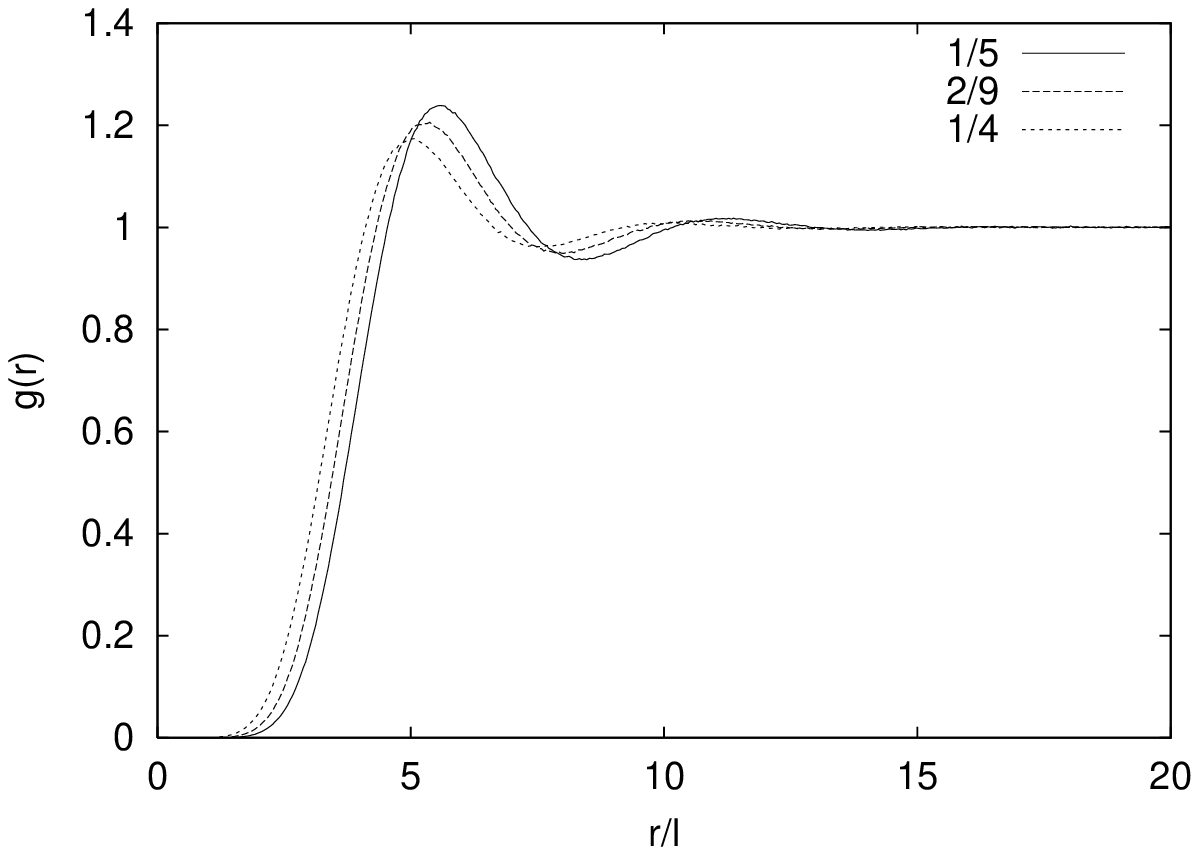}
\end{center}
\caption[]{Pair distribution function for the anyon wave function
           at filling factors $\nu=1/3,2/5,1/2$
            and $1/5,2/9,1/4$ obtained after a MC simulation
            in disk geometry for $N=196$ particles. }
%\label{g3standard}
\label{g35standard}
\end{figure}

{\it Discussions.}$-$We are now in the position to discuss
the exchange corrections. To this end, it is instructive to 
focus on the limiting case of $\nu=1/2$. 
According to the neutral fermion theory~\cite{dhlee},
at filling of $\nu =1 /2$, a liquid of
$\pm 1/2$ charged anyons (neutral fermions) floats
on top of a Bose quantum Hall liquid. In view that the Bose
quantum Hall liquid is described by Eq. (11) with $n=\infty$, 
the energy for the dipoles (neutral fermions) is nothing but
$\Delta_1$. 
The neutral fermions and the Bose quantum Hall fluid are decoupled in that
the neutral fermions experience no magnetic field and their contribution
to $\sigma_{xy}$ is 0. The decoupling of the neutral fermions and the 
Bose quantum fluid has
important consequences for the low-energy physics. 
In general, the wave functions
factorize into products and the corresponding anyon excitations and 
neutral fermion
excitations have different behavior.   
 
It is tempting to develop a 
perturbation for the exchange corrections. 
However, the ``expansion parameter'', $|1-1/n|$, is not small. 
Analogous to the case of $\nu=1/2$,
it is reasonable to view CFs as a combination of anyons and 
dipole interactions. The CFs and anyons experience an effective
magnetic field corresponding to $n$-filled Landau bands.   
The ``gradient correction'' to the 
anyon wave function can be identified as a 
dipole field that experiences no residual magnetic field. 
A fluid of dipoles is expected to be compressible~\cite{dhlee}, which decouples from the 
incompressible quantum fluid described by anyons due to different symmetry 
properties under ${\cal T}$. 
In fact,
our results from Eq. (12) suggest an effect of vortex-charge separation for
the CFs. Naturally, the effect of the dipole field 
is expected to be proportional to $(1-1/n)$.
Further quantitative studies are
clearly desirable.  

%
%  new addition
%
In summary, we introduced and studied the properties of a specific
microscopic anyon wave function for the FQHE.
Our MC results based on this anyon wave function
provide a {\it quantitative} measure of the contributions from the anyon
picture of the FQHE, which was discussed {\it qualitatively} by
Ma and Zhang.~\cite{MaZhang} 
The comparison with Jain's CF results reveals interesting exchange
effects arising from the approximations involved in the 
anyon wave function.

\acknowledgments
The work of XQW was supported by
US Army Grant No. DAAE07-99-C-L065, and
NASA Grant No. NAG3-2833,
National Science Foundation Grant No. DMR-02-05328.
Part of the work of OC was supported 
by the Office of the Vice-President
for Research and Development of Prairie View A\&M University through
a 2003-2004 Research Enhancement Program grant.

\end{document}